\documentclass[11pt]{article}
\usepackage{epsfig}
\renewcommand{\arraystretch}{1.3}

\catcode`\@=11
\def\marginnote#1{}

\newcount\hour
\newcount\minute
\newtoks\amorpm
\hour=\time\divide\hour by60
\minute=\time{\multiply\hour by60 \global\advance\minute by-\hour}
\edef\standardtime{{\ifnum\hour<12 \global\amorpm={am}%
        \else\global\amorpm={pm}\advance\hour by-12 \fi
        \ifnum\hour=0 \hour=12 \fi
        \number\hour:\ifnum\minute<10 0\fi\number\minute\the\amorpm}}
\edef\militarytime{\number\hour:\ifnum\minute<10 0\fi\number\minute}

\def\draftlabel#1{{\@bsphack\if@filesw {\let\thepage\relax
      \xdef\@gtempa{\write\@auxout{\string
          \newlabel{#1}{{\@currentlabel}{\thepage}}}}}\@gtempa \if@nobreak
    \ifvmode\nobreak\fi\fi\fi\@esphack} \gdef\@eqnlabel{#1}}
    \def\@eqnlabel{}
\def\@vacuum{}
\def\draftmarginnote#1{\marginpar{\raggedright\scriptsize\tt#1}}

\def\draft{
  \oddsidemargin -.5truein
  \def\@oddfoot{\footnotesize \sl preliminary draft \hfil
    \rm\thepage\hfil\sl\today\quad\militarytime}
  \let\@evenfoot\@oddfoot \overfullrule 3pt
    \let\label=\draftlabel
    \let\marginnote=\draftmarginnote
  \def\@eqnnum{(\theequation)\rlap{\kern\marginparsep\tt\@eqnlabel}%
    \global\let\@eqnlabel\@vacuum}

  }

\makeatletter
\newdimen\normalarrayskip              
\newdimen\minarrayskip                 
\normalarrayskip\baselineskip
\minarrayskip\jot
\newif\ifold             \oldtrue            \def\new{\oldfalse}
\def\arraymode{\ifold\relax\else\displaystyle\fi} 
\def\eqnumphantom{\phantom{(\theequation)}}     
\def\@arrayskip{\ifold\baselineskip\z@\lineskip\z@
     \else
     \baselineskip\minarrayskip\lineskip2\minarrayskip\fi}
\def\@arrayclassz{\ifcase \@lastchclass \@acolampacol \or
\@ampacol \or \or \or \@addamp \or
   \@acolampacol \or \@firstampfalse \@acol \fi
\edef\@preamble{\@preamble
  \ifcase \@chnum
     \hfil$\relax\arraymode\@sharp$\hfil
     \or $\relax\arraymode\@sharp$\hfil
     \or \hfil$\relax\arraymode\@sharp$\fi}}
\def\@array[#1]#2{\setbox\@arstrutbox=\hbox{\vrule
     height\arraystretch \ht\strutbox
     depth\arraystretch \dp\strutbox
     width\z@}\@mkpream{#2}\edef\@preamble{\halign
\noexpand\@halignto
\bgroup \tabskip\z@ \@arstrut \@preamble \tabskip\z@ \cr}%
\let\@startpbox\@@startpbox \let\@endpbox\@@endpbox
  \if #1t\vtop \else \if#1b\vbox \else \vcenter \fi\fi
  \bgroup \let\par\relax
  \let\@sharp##\let\protect\relax
  \@arrayskip\@preamble}
\def\eqnarray{\stepcounter{equation}%
              \let\@currentlabel=\theequation
              \global\@eqnswtrue
              \global\@eqcnt\z@
              \tabskip\@centering
              \let\\=\@eqncr
 \halign to \displaywidth\bgroup
    \eqnumphantom\@eqnsel\hskip\@centering
    $\displaystyle \tabskip\z@ {##}$%
    \global\@eqcnt\@ne \hskip 2\arraycolsep
         $\displaystyle\arraymode{##}$\hfil
    \global\@eqcnt\tw@ \hskip 2\arraycolsep
         $\displaystyle\tabskip\z@{##}$\hfil
         \tabskip\@centering
    &{##}\tabskip\z@\cr}
\begingroup\ifx\undefined\newsymbol \else\def\input#1 {\endgroup}\fi
\newfont{\hr}{msbm10}
\newfont{\ams}{msam10}

\voffset= - 1.2in
\hoffset= - 1.0in         

\def\beq{\begin{equation}}
\def\eeq{\end{equation}}
\def\ba{\beq\new\begin{array}{c}}
\def\ea{\end{array}\eeq}
\def\be{\ba}
\def\ee{\ea}
\def\stackreb#1#2{\mathrel{\mathop{#2}\limits_{#1}}}

\def\res{{\rm res}}

\def\F{{\cal F}}

\def\d{\partial}
\def\N2{${\cal N}=2$}

\def\1N{${\cal N}=1$}
\def\4N{${\cal N}=4$}

\title{{\bf
DV and WDVV}
\vspace{.5cm}}
\author{{\bf L. Chekhov}\thanks{E-mail: \ chekhov@mi.ras.ru}
\date{ } \\ {\small
{\it Steklov Mathematics Institute, Moscow, Russia}}\\ \\
{\bf A. Marshakov}\thanks{E-mail: \ mars@lpi.ac.ru
}\ \ \ {\bf A. Mironov}\thanks{E-mail:
\ mironov@itep.ru; mironov@lpi.ac.ru}
\date{ } \\
{\small {\it Theory Department, Lebedev Physics Institute}
and {\it ITEP, Moscow, Russia}}\\ \\
{\bf D.Vasiliev}\thanks{E-mail: \ dmtrvass@gate.itep.ru}
\date{ } \\ {\small {\it MIPT} and
{\it ITEP, Moscow, Russia}}}

\begin{document}

\setcounter{footnote}{3}

\maketitle

\vspace{-10cm}

\begin{center}
\hfill FIAN/TD-02/03\\
\hfill ITEP/TH-04/03\\
\end{center}

\vspace{6.5cm}

\begin{abstract}
We prove that the quasiclassical tau-function of the multi-support solutions
to matrix models, proposed recently by Dijkgraaf and Vafa to be related to
the Cachazo-Intrilligator-Vafa
superpotentials of the ${\cal N}=1$ supersymmetric Yang-Mills theories,
satisfies the Witten-Dijkgraaf-Verlinde-Verlinde equations.
\end{abstract}

\vspace{1.5cm}

\setcounter{footnote}{0}
\section{Introduction}

The Witten-Dijkgraaf-Verlinde-Verlinde (WDVV) equations \cite{WDVV} in the most
general form can be written \cite{MMM} as system of algebraic relations
\be
\label{WDVV}
\F_I{\F}_J^{-1}\F_K = \F_K{\F}_J^{-1}\F_I, \ \ \ \ \ \ \forall\ I,J,K
\ee
for the third derivatives
\be
\label{matrF}
\|{\F}_{I}\|_{JK}=
{\d^3\F\over\d T_I\,\d T_J\,\d T_K} \equiv\F_{IJK}
\ee
of some function $\F ({\bf T})$. Have been appeared first in the context of
topological string theories \cite{WDVV}, they were rediscovered later
on in much larger class of physical theories where the exact answer
for a multidimensional theory could be
expressed through a single holomorphic function of several complex variables
\cite{MMM,forms,MMMlong,Veselov,Luuk,BMRWZ,WDVVmore}.

Recently, a new example of similar relations between the superpotentials
of ${\cal N}=1$ supersymmetric gauge theories in four dimensions and free
energies of matrix models in the planar limit was proposed \cite{CIV,DV}.
It has been realized that superpotentials in some ${\cal N}=1$
four-dimensional Yang-Mills theories can be expressed through a
single holomorphic function \cite{CIV} that can be further identified with free
energy of the multi-support solutions to matrix models in the planar
limit \cite{DV}.
A natural question which immediately arises in this context is whether these
functions -- the quasiclassical tau-functions, determined by
multi-support solutions to matrix models, satisfy the WDVV equations? In
the case of positive answer this is rather important, since multi-support
solutions to the matrix models can play a role of ``bridge" between
topological string theories and Seiberg-Witten theories \cite{SW} which
give rise to two different classes of solutions to the WDVV equations
(see, e.g., \cite{Dub} and \cite{MirWDVV,Mtmf}).

This question was already addressed in \cite{imo}, where it was shown that
the multicut solution to one-matrix model satisfies the WDVV equations. However,
this was verified only {\em perturbatively} and, what is even more important,
for a particular {\em non-canonical!} (and rather strange) choice of variables.

In this paper, we demonstrate that the quasiclassical
tau-function of the multi-support solution satisfies the WDVV equations as a
function of {\em canonical} variables identified with the periods and residues
of the generating meromorphic one-form $dS$ \cite{KriW}. An exact proof of
this statement consists of two steps.
The first, most difficult step is to find the residue formula for the third
derivatives (\ref{matrF}) of the matrix model free energy. Then, using an
associative algebra, we immediately prove that free energy of multi-support
solution satisfies WDVV equations, upon the number of independent variables
is fixed to be equal to the number of critical points in the residue formula.

In sect.~\ref{ss:taumatr}, we define
the free energy of the multi-support matrix model in terms of the
quasiclassical tau-function \cite{KriW} along the line of
\cite{David,CM,KM}. In sect.~\ref{ss:residue}, we derive the residue formula
for the third derivatives of the quasiclassical tau-function for the
variables associated both with the periods $\{ {\bf S}\} = \{ S_i\}$ and residues
$\{ {\bf t}\} = \{
t_i\}$ of the generating differential $dS$. In sect.~\ref{ss:proof}, we prove
that the free energy of the multisupport solution $\F ({\bf T})$
solves the WDVV equations (\ref{WDVV}) as a function of the full set of
variables $\{ {\bf T}\} = \{ {\bf S}, {\bf t}\}$ whose total number should
be fixed to be equal to the number of critical points in the residue
formula for the third derivatives (\ref{matrF}). In sect.~\ref{ss:expl}, we
verify this statement explicitly for the first nontrivial
case where the total number of variables is equal to four.
\footnote{Let us point out that the WDVV equations (\ref{WDVV}) are nontrivial
only for the functions of at least three independent variables. However,
as we see below, the structure of residue formula for the matrix model free
energy requires the minimal number of independent variables to be at least
four! From this point of view, the origin of the
``experimental observation" of \cite{imo} valid for a
function of three variables still remains unclear to us.}
Finally we present several concluding remarks and discuss possible
generalizations.

We restrict ourselves by the ${\cal N}=1$ supersymmetric theories
without flavours originally considered in \cite{CIV}. The results can be
easily generalized. Note that the literature on the subject is already
quite extensive \cite{DV1,DV2}, and different interesting developments of
the issues discussed in this paper can be immediately obtained.

\section{Tau-function of multi-support matrix model
\label{ss:taumatr}}

We first exactly define what we call below the multi-support
free energy of the matrix model in the planar limit. We are
mostly doing with one-matrix integrals
\footnote{The generalization to the two-matrix case \cite{KM} is rather
straightforward and will be discussed in the last section.}
of the form
\be
\label{mamo}
Z=\int\ d\Phi\ e^{{1\over\hbar} {\rm Tr} W(\Phi)}
\ee
where the potential $W(\Phi)$ is supposed to be a polynomial of a degree
$(n+1)$. The free energy in the planar
limit of (\ref{mamo}) can be defined as the first term in the expansion
\be
\label{planar}
F({\bf t},t_0) = \lim\left( \log Z\left({\bf t}\right)\right)
= \sum_{g=0}^{\infty} \hbar^{2g-2} F_g({\bf t},t_0)
\ee
implying $N\to\infty$, $\hbar\to 0$ with $N\hbar = t_0$ being fixed.
In what follows we are only interested in the first term of this expansion,
$F_0({\bf t},t_0)$. In fact, we deal with another quantity,
$\F ({\bf t},t_0,{\bf S})$, where $S_i = \hbar N_i$, $\sum S_i = t_0$ are {\em
extra} variables -- the filling numbers of (metastable) vacua. In order to
get from this quantity $F_0({\bf t},t_0)$, one needs to minimize the free energy
with respect to the filling numbers, ${\partial \F\over\partial S_i}=0$.
However, one can still preserve $S_i$ as free parameters introducing the
"chemical potentials".

\begin{figure}[tb]
\epsfysize=6cm
\centerline{\epsfbox{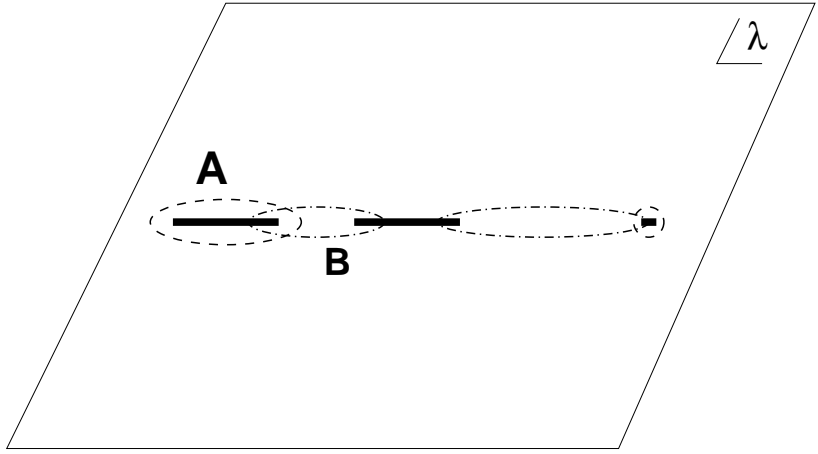}}
\caption{Cuts in the $\lambda$- or ``eigenvalue" plane for the
planar limit of 1-matrix model. The eigenvalues are supposed to be
located ``on" the cuts. The distribution of eigenvalues is governed by the
period integrals ${\bf S} = \oint_{\bf A} \rho(\lambda)d\lambda$ along the
corresponding cycles
and the dependence of partition function on ``distributions" $S_i$ is
given by the quasiclassical tau-function ${\d{\cal F}\over\d {\bf S}} =
\oint_{\bf B} \rho(\lambda)d\lambda$.
}
\label{fi:cuts}
\end{figure}

The origin of the new variables $S_i$ becomes rather transparent after one says
that instead of direct computation of (\ref{mamo}) this problem is replaced
by the saddle point approximation -- finding the extremum of the
functional $F_0 \left[\rho(\lambda)\right] \propto \int W\rho -
\int\int \rho (\lambda_1)\log\left|\lambda_1-\lambda_2\right|\rho (\lambda_2)\ +
\Pi_0 \left(\int\rho - t_0\right)$
where $\Pi_0$ is just a Lagrange multiplier
to fix the total normalization of the eigenvalue density.
This latter condition means the saddle point
equation is non-trivial only on the support of $\rho$. For one matrix model, this
support can be presented as a set $\{D_i\}$ of cuts in complex eigenvalue plane,
see fig.~\ref{fi:cuts}.
Then, one should add to this functional
the term $\sum \Pi_i \left(\int_{D_i}\rho - S_i\right)$, which
via Lagrange multipliers, controls the filling numbers at each cut, i.e.
to consider
\be
\label{variF}
\F \left({\bf t},t_0,{\bf S}\right) \propto \int W\rho -
\int\int \rho (\lambda_1)\log\left|\lambda_1-\lambda_2\right|\rho (\lambda_2)\ +
\Pi_0 \left(\int\rho - t_0\right)
+ {\sum}' \Pi_i \left(\int_{D_i}\rho - S_i\right)
\ee
Then, the extra variables appear due to the extra information
hidden in (\ref{variF}) compare to (\ref{mamo}) -- the structure of
nontrivial eigenvalue supports.
It is well-known that at ``critical" densities ${\delta\F\over\delta\rho} =
0$,
(\ref{variF}) is
a (logarithm) of quasiclassical tau-function \cite{KriW} (see, e.g.,
\cite{CM,KM}).

In principle, in order to compare $\F$ with the matrix
model quantity $F_0$, one needs to put further restrictions to get rid of
the metastable vacua. This would lead to shrinking part of the cuts into
the double points (see discussions of
these issues, say, in \cite{CM}). Here we would forget this issue and
consider smooth curves (\ref{dvc}) with only two marked points at
infinities on two $\lambda$-sheets of the curve (\ref{dvc}).

Note
also that in (\ref{variF}) one can make two different
natural choices for the set of
new independent variables: the first choice corresponds to independent
filling of all cuts, then $t_0 = \sum S_i$, while the second choice
corresponds to choosing as independent $t_0$ and all $S_i$ except of
corresponding to one of the cuts (that is why the corresponding sum
in (\ref{variF}) is denoted as ${\sum}'$). These choices are related by
the linear change
of variables which does not influence the WDVV equations (see
\cite{MMM,dWM}). The first choice is more ``symmetric" while the second one
corresponds to the canonical choice of variables in the sense of \cite{KriW}
or to the homology basis on smooth curve (\ref{dvc}) with added marked
points at two infinities. We use both of them below
depending on convenience.


The complex curve of one-matrix model ``comes from" the loop equations
(see, for example, \cite{Migdal}) and can be written in the form
\be
\label{dvc}
y^2 = W'(\lambda)^2 + f(\lambda) \equiv R(\lambda)
\ee
with the matrix model potential (\ref{mamo}) parameterized as
\be
\label{mmpot}
W(\lambda) = \sum_{l\geq 0} t_{l+1} \lambda^{l+1}
\ee
or
\be
\label{mmpot'}
W'(\lambda) = \sum_{l=0}^n (l+1)t_{l+1} \lambda^l
\ee
being the polynomial of $n$-th degree in our conventions. The coefficients
of the function
\be
\label{f}
f(\lambda) = \sum_{k=0}^{n-1}f_k\lambda^k
\ee
are related to the extra data (the filling numbers) of the multicut
solution. The eigenvalue density $\rho(\lambda)$ is the imaginary part of
$y(\lambda)$ and vanishes outside the cuts. Therefore, the eigenvalue
distribution ${\bf S}$ can be fixed by the periods
\be
\label{DVper}
S_i = \oint_{A_i}dS
\ee
of the generating differential
\be
\label{dvds}
dS = yd\lambda
\ee
taken around the eigenvalue supports to be identified (except for one
of the supports) with the canonical ${\bf A}$-cycles. Then
\be
\label{canoDV}
{\d dS\over\d S_i} = d\omega_i
\\
\oint_{A_i}d\omega_j = \delta_{ij}
\ee
when the derivatives are taken at fixed coefficients $\{ t_l\}$
of the potential (\ref{mmpot'}). One can show that the Lagrangian
multipliers in (\ref{variF}) are given by integrals of the same
generating differential (\ref{dvds}) over the dual contours (see
fig.~\ref{fi:cuts})
\be
\label{DVF0}
\Pi _i = \oint_{B_i}dS
\ee
To the set of parameters (\ref{DVper}) one should also add
\footnote{By the $\infty$-point in what follows we call for short the point
$\infty_+$ or $\lambda=\infty$ on the ``upper" sheet of hyperelliptic Riemann
surface (\ref{dvc}) corresponding to the positive sign of the square root,
i.e. to $y=+\sqrt{W'(\lambda)^2+f(\lambda)}$.} the
total number of eigenvalues $N\hbar = t_0$
\be
\label{t0}
\res_\infty\left(dS \right) = {f_{n-1}\over 2(n+1)t_{n+1}} \equiv t_0
\ee
and the parameters of the potential (\ref{mmpot}), (\ref{mmpot'}), which
can be equivalently written as
\be
\label{LGtimes}
t_k = {1\over k}\res_\infty\left(\lambda^{-k}dS\right)
\\
k=1,\dots,n
\ee
while the leading term coefficient $t_{n+1}$ is supposed to be fixed
(we will discuss this issue in detail below). Then
\be
\label{bipole}
d\Omega_0 = {\d dS \over \d t_0} =(n+1)t_{n+1}{\lambda^{n-1} d\lambda\over y} +
{1\over 2}\sum_{k=0}^{n-2}{\d f_k\over\d t_0}{\lambda^k d\lambda\over y}
\ee
and the dependence of $\{ f_k\}$ with $k=0,1,\dots,n-2$ on $t_0$ is fixed
by the condition
\be
\oint_{A_i}\left((n+1)t_{n+1}{\lambda^{n-1} d\lambda\over y} +
\sum_{k=0}^{n-2}{\d f_k\over\d t_0}{\lambda^k d\lambda\over y}\right)=0
\ee
which for $i=1,\dots,n-1$ gives exactly $n-1$ relations on the derivatives
of $f_0,f_1,\dots,f_{n-2}$ w.r.t. $t_0$.
The bipole differential (\ref{bipole}) can be also presented as
\be
\label{bp}
d\Omega_{0} = d\log
\left({E(P,\infty)\over E(P,\infty_-)}\right)
\ee
where $E(P,P')$ is the Prime form. Differential (\ref{bp}) obviously obey
the properties
\be
\res_{\infty}d\Omega_{0} = - \res_{\infty_-}d\Omega_{0} = 1
\\
\oint_{A_i}d\Omega_{0} = 0,\ \ \ \ \ \ i=1,\dots,n-1
\ee
For the derivatives w.r.t. parameters of the potential
(\ref{LGtimes}), one gets
\be
\label{omes}
d\Omega_k = {\d dS \over \d t_k} =
{W'(\lambda)k\lambda^{k-1} d\lambda\over y} +
{1\over 2}\sum_{j=0}^{n-2}{\d f_j\over\d t_k}{\lambda^k d\lambda\over y}
\ee
obeying
\be
\label{peom}
\oint_{A_i}d\Omega_k =
\oint_{A_i}{W'(\lambda)k\lambda^{k-1} d\lambda\over y} +
{1\over 2}\sum_{j=0}^{n-2}{\d f_j\over\d t_k}\oint_{A_i}
{\lambda^k d\lambda\over y} = 0
\ee
and this is again a system of linear equations on ${\d f_j\over\d t_k}$.
To complete the setup
one should also add to (\ref{DVF0})
the following formulas
\footnote{\label{reg}
Naively understood the integral in (\ref{dfdt0}) is divergent
and should be supplemented by some proper regularization.
In what follows we ignore this
subtlety since it does not influence the residue formulas for the third
derivatives, those one really needs for the WDVV equations (\ref{WDVV}). The
simplest way to avoid these complications is to think of the pair of marked
points $\infty$ and $\infty_-$ as of degenerate handle; then the residue
(\ref{t0}) comes from degeneration of the extra $A$-period, while the
integral (\ref{dfdt0}) from degeneration of the extra $B$-period.}:
\be
\label{dfdt0}
\Pi_0 = \int_{\infty_-}^{\infty_+}dS
\ee
(we again remind that, instead of $t_0$, the parameter $S_n =
t_0 - \sum_{i=1}^{n-1}S_i $ can be used equivalently) and
\be\label{v}
v_k = \res_\infty\left(\lambda^k dS\right), \ \ \ \ k>0
\ee

\begin{figure}[tp]
\vspace*{-2.5cm}
\epsfysize=11cm
\centerline{\epsfbox{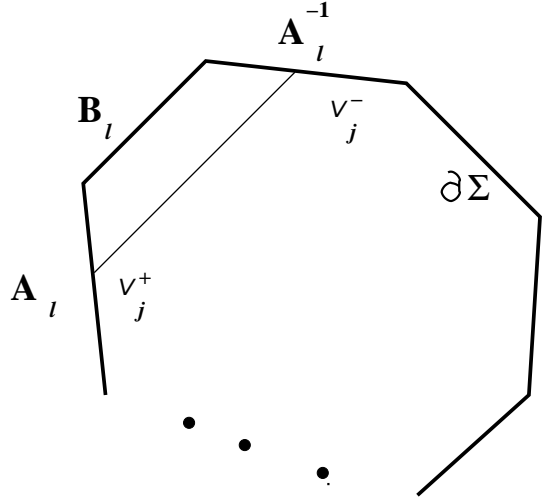}}
\vspace*{-1.5cm}
\caption{Cut Riemann surface with boundary $\d\Sigma$. The integral over the
boundary can be divided into several pieces (see formula (\ref{derdpm})). In
the process of computation we use the fact that the boundary values of
Abelian integrals
$v_j^\pm$ on two copies of the cut differ by period integral of
the corresponding differential
$d\omega_j$ over the dual cycle.}
\label{fi:cut}
\end{figure}

On genus $g=n-1$ smooth Riemann surface (\ref{dvc}), there
are $2g=2n-2$ independent noncontractable contours
which can be split into the so-called
${\bf A}\equiv\{ A_i\}$ and ${\bf B}\equiv\{ B_i\}$,
$i = 1,\dots,g$, cycles with the
intersection form $A_i\circ B_j = \delta _{ij}$.
The canonical holomorphic differentials (\ref{canoDV})
are normalized to the ${\bf A}$-cycles,
and their integrals along the ${\bf B}$-cycles give the period matrix,
\be\label{pemat}
\oint _{B_j}d\omega _i =  T_{ij}
\ee
To check integrability of (\ref{DVF0}) and (\ref{dfdt0}) one needs to verify
the symmetricity of the second derivatives. For the part related with the
derivatives only w.r.t. the variables (\ref{DVper}), this is just a
symmetricity of the period matrix of (\ref{dvc}) and follows
from the Riemann bilinear relations for the canonical holomorphic
differentials (\ref{canoDV})
\be
\label{sypema}
0=\int_{\Sigma} d\omega_i\wedge d\omega_j=
\sum_k\left( \oint_{A_k}d\omega_i\oint_{B_k}
d\omega_j-\oint_{A_k}
d\omega_j\oint_{B_k}d\omega_i \right)
= T_{ij} - T_{ji}
\ee
Analogously
\be
\label{syToda}
0=\int_{\Sigma} d\omega_i\wedge d\Omega_0=
\sum_k\left( \oint_{A_k}d\omega_i\oint_{B_k}
d\Omega_0-\oint_{A_k}
d\Omega_0\oint_{B_k}d\omega_i \right) + \\ +
\res_\infty(d\omega_i)\int_{\infty_-}^{\infty_+}d\Omega_0
- \res_\infty(d\Omega_0)\int_{\infty_-}^{\infty_+}d\omega_i
= \oint_{B_i}d\Omega_0 - \int_{\infty_-}^{\infty_+}d\omega_i
\ee
Formula (\ref{sypema}) means that
\be
\label{intDV1}
{\d\Pi_i\over\d S_j} = T_{ji} = T_{ij} = {\d\Pi_j\over\d S_i}
\ee
while from (\ref{syToda}) one gets
\be
\label{intDV2}
{\d\Pi_j\over\d t_0} = {\d\Pi_0\over\d S_j}
\ee
This allows one to introduce the function
$\F({\bf T}) = \F ({\bf S},t_0,{\bf t})$ such that
\be
\label{FDV}
{\d \F \over\d S_j} = \Pi_j, \ \ \ \
{\d \F \over\d t_0} = \Pi_0, \ \ \ \
{\d \F \over\d t_k} = v_k
\ee
The integrability of the last relation can be checked similar to
(\ref{intDV1}), (\ref{intDV2}) with the help of Riemann bilinear relations
involving the Abelian integrals $\Omega_k = \int^P d\Omega_k$,
for example (cf. e.g. with \cite{RG}, where similar relations were used for the
quasiclassical tau-function of the Seiberg-Witten theory):
\be
\res_\infty\left(\Omega_kd\omega_i\right) =
\oint_{\d\Sigma}\Omega_kd\omega_i =
\sum_l\left(\int_{A_l}\Omega_k^+d\omega_i - \int_{A_l}\Omega_k^-
d\omega_i\right) -
\sum_l\left(\int_{B_l}\Omega_k^+d\omega_i - \int_{B_l}\Omega_k^-
d\omega_i\right)
=
\\
= \sum_l \left(\oint_{B_l}d\Omega_k\oint_{A_l}d\omega_i -
\oint_{A_l}d\Omega_k\oint_{B_l}d\omega_i
\right) = \oint_{B_i}d\Omega_k
\ee
where $\d\Sigma$ is the cut Riemann surface (\ref{dvc}) (see
fig.~\ref{fi:cut}), and in the last equality we used (\ref{peom}).

\section{Residue formula
\label{ss:residue}}

\subsection{Holomorphic differentials
\label{ss:reshol}}

Let us now derive the formulas for the third derivatives of $\F$,
following the way proposed by I.Krichever \cite{KriW,Kripri}.
We first note that the derivatives of the elements
of period matrix (in this section, for simplicity, we set the
coefficients of potential (\ref{mmpot'}) to be fixed)
can be expressed through the integral over the ``boundary"
$\d\Sigma$ of cut Riemann surface $\Sigma$ (see fig.~\ref{fi:cut})
\be
\label{dpm}
{\d T_{ij}\over \d S_k}\equiv \d_k T_{ij} = \int_{B_j}\d_k d\omega_i =
\int_{\d\Sigma}\omega_j\d_k d\omega_i
\ee
where $\omega_j = \int^P d\omega_j$ are the Abelian integrals, whose values
on two copies of cycles on the cut Riemann surface (see also fig.~\ref{fi:cut})
are denoted below as $\omega_j^\pm$.
Indeed, the computation of the r.h.s. of (\ref{dpm}) gives
\be
\label{derdpm}
\int_{\d\Sigma}\omega_j\d_k d\omega_i =
\sum_l\left(\int_{B_l}\omega_j^+\d_kd\omega_i -
\int_{B_l}\omega_j^-\d_kd\omega_i\right) -
\sum_l\left(\int_{A_l}\omega_j^+\d_kd\omega_i -
\int_{A_l}\omega_j^-\d_kd\omega_i\right) = \\ =
\sum_l\oint_{B_l}\left(\oint_{A_l}d\omega_j\right)\d_kd\omega_i  -
\sum_l\oint_{A_l}\left(\oint_{B_l}d\omega_j\right)\d_kd\omega_i =
\\ =
\sum_l\left(\oint_{A_l}d\omega_j\right)\oint_{B_l}\d_kd\omega_i  -
\sum_l\left(\oint_{B_l}d\omega_j\right)\oint_{A_l}\d_kd\omega_i\
\stackreb{\oint_{A_i}d\omega_j=\delta_{ij}}{=}\ \d_k T_{ij}
\ee
One can now rewrite (\ref{dpm}) as
\be
\label{dpmres}
\d_k T_{ij} = \int_{\d\Sigma}\omega_j\d_k d\omega_i =-
\int_{\d\Sigma}\d_k \omega_j d\omega_i = \sum\res_{d\lambda = 0}
\left(\d_k \omega_j d\omega_i\right)
\ee
where the sum is taken over all residues of the integrand, i.e.
over all residues of $\d_k \omega_j$ since $d\omega_i$ are
holomorphic. In order to investigate these singularities and
understand the last equality in (\ref{dpmres}), we
discuss first how to take derivatives $\d_k$ w.r.t. moduli, or
introduce the corresponding connection.

To this end, we introduce a
covariantly constant function -- the hyperelliptic co-ordinate $\lambda$,
i.e. such a connection that $\d_k\lambda=0$. Roughly
speaking, the role of covariantly constant function can be played
by one of two co-ordinates --
in the simplest possible description of complex curve by a
single equation on two complex variables. Then, using this
equation, one may express the other co-ordinate as a function of
$\lambda$ and moduli. Any Abelian integral $\omega_j$
can be then expressed in terms of $\lambda$,
and in the vicinity of critical points $\{\lambda_\alpha\}$
where $d\lambda=0$
(for a general (non-singular) curve this is always true) we get an expansion
\be
\label{vexp}
\omega_j(\lambda)\ \stackreb{\lambda\to\lambda_\alpha}{=}\ \omega_{j\alpha} +
c_{j\alpha}\sqrt{\lambda-\lambda_\alpha}+ \dots
\ee
whose derivative
\be
\label{dervj}
\d_k\omega_j \equiv \left.\d_k\omega_j\right|_{\lambda= {\rm const}} =
- {c_{j\alpha}\over
2\sqrt{\lambda-\lambda_\alpha}}\d_k\lambda_\alpha + {\rm regular}
\ee
gives first order poles at $\lambda=\lambda_\alpha$ up to
regular terms which do not contribute to
(\ref{dpmres}). The exact coefficient in (\ref{dervj}) can be
computed for $\lambda$ related with
the generating differential $dS = yd\lambda$. Then, using
\be
\label{yl}
y(\lambda)\ \stackreb{\lambda\to\lambda_\alpha}{=}\
\Gamma_\alpha\sqrt{\lambda-\lambda_\alpha}+ \dots
\ee
where $\Gamma_\alpha =
\sqrt{\prod_{\beta\neq\alpha}(\lambda_\alpha-\lambda_\beta)}$ or
\be
\label{deryl}
{\d\over\d t_k} y(\lambda) = - {\Gamma_\alpha\over
2\sqrt{\lambda-\lambda_\alpha}}{\d\lambda_\alpha\over\d t_k} + {\rm regular}
\ee
together with
\be
\label{difyv}
dy = {\Gamma_\alpha \over
2\sqrt{\lambda-\lambda_\alpha}}\ d\lambda + {\rm regular}
\ee
and
\be
\label{difzv}
d\omega_j = {c_{j\alpha}\over
2\sqrt{\lambda-\lambda_\alpha}}\ d\lambda + \dots
\ee
and, following from (\ref{dvds}) and (\ref{canoDV})
\be
d\omega_k = \d_k dS = - {\Gamma_\alpha\d_k\lambda_\alpha\over
2\sqrt{\lambda-\lambda_\alpha}}\ d\lambda + {\rm regular}
\ee
one finally gets for (\ref{dpmres})
\be
\sum\res\left(\d_k \omega_j d\omega_i\right) = \sum_\alpha
\res\left( {c_{j\alpha}\d_k\lambda_\alpha\over
2\sqrt{\lambda-\lambda_\alpha}}d\omega_i\right)
= \sum_\alpha
\res\left( {d\omega_j\over
d\lambda}d\omega_i\d_k\lambda_\alpha\right)=
\sum_\alpha
\res\left( {d\omega_id\omega_jd\omega_k\over
d\lambda dy}\right)
\ee
In hyperelliptic situation, the derivation presented above is
equivalent to using the Fay formula \cite{Fay}
\be
\label{Fay}
\frac{\partial
T_{ij}}{\partial \lambda_\alpha} = \hat
\omega_i(\lambda_\alpha)\hat\omega_j(\lambda_\alpha)
\label{derram}
\ee
where $\hat\omega_i(\lambda_\alpha) =
\left.{d\omega_i(\lambda)\over
d\sqrt{\lambda-\lambda_\alpha}}\right|_{\lambda=\lambda_\alpha}$ is
the "value" of canonical differential at a critical point.

\subsection{Meromorphic differentials
\label{ss:resmero}}

Almost in the same way the residue formula can be derived for the
meromorphic differentials (\ref{omes}).
One gets
\be
{\d\F \over\d t_k} = \res_\infty \left(\lambda^k dS \right), \ \ \ \ k>0
\ee
therefore
\be
{\d^2\F \over\d t_k\d t_n} =
\res_\infty \left(\lambda^k d\Omega_n\right) =
\res_\infty \left((\Omega_k)_+ d\Omega_n\right)
\ee
where $\left(\Omega_k\right)_+$ is the singular part of the integral of
1-form $d\Omega_k$.
Further
\be
{\d\over\d t_m}\res_\infty \left(\lambda^k d\Omega_n\right) =
\res_\infty \left(\lambda^k {\d d\Omega_n\over\d t_m}\right) =
 -\res_\infty \left((d\Omega_k)_+ {\d \Omega_n\over \d t_m}\right) =
-\res_\infty \left(d\Omega_k {\d \Omega_n\over \d t_m}\right)
\ee
The last expression can be rewritten as
\be
\label{3mero}
-\res_\infty \left(d\Omega_k {\d \Omega_n\over \d t_m}\right) =
\oint_{\d\Sigma}\left(d\Omega_k {\d \Omega_n\over \d t_m}\right) +
\sum \res_{\lambda_\alpha} \left(d\Omega_k {\d \Omega_n\over \d t_m}\right)
= \sum \res_{\lambda_\alpha} \left(d\Omega_k {\d \Omega_n\over \d t_m}\right)
\ee
since $\oint_{\d\Sigma}\left(d\Omega_k {\d \Omega_n\over \d t_m}\right) =0$
due to $\oint_{A_i}d\Omega_n = 0$, (cf. with (\ref{derdpm})):
\be
\label{derome}
\int_{\d\Sigma}\Omega_j{\d\over\d t_k} d\Omega_i =
\sum_l\left(\int_{B_l}\Omega_j^+{\d\over\d t_k}d\Omega_i -
\int_{B_l}\Omega_j^-{\d\over\d t_k}d\Omega_i\right) -
\sum_l\left(\int_{A_l}\Omega_j^+{\d\over\d t_k}d\Omega_i -
\int_{A_l}\Omega_j^-{\d\over\d t_k}d\Omega_i\right) = \\ =
\sum_l\oint_{B_l}\left(\oint_{A_l}d\Omega_j\right){\d\over\d t_k}d\Omega_i
-
\sum_l\oint_{A_l}\left(\oint_{B_l}d\Omega_j\right){\d\over\d t_k}d\Omega_i =
\\ =
\sum_l\left(\oint_{A_l}d\Omega_j\right)\oint_{B_l}{\d\over\d t_k}d\Omega_i
-
\sum_l\left(\oint_{B_l}d\Omega_j\right)\oint_{A_l}{\d\over\d t_k}d\Omega_i\
\stackreb{\oint_{A_i}d\Omega_j=0}{=}\ 0
\ee
Now, as in the holomorphic case one takes
\be
\label{oexp}
\Omega_n(\lambda)\ \stackreb{\lambda\to\lambda_\alpha}{=}\
\Omega_{n\alpha} +
\gamma_{n\alpha}\sqrt{\lambda-\lambda_\alpha}+ \dots
\ee
and, therefore
\be
\label{deroj}
{\d\over\d t_k}\Omega_j \equiv \left.{\d\over\d t_k}\Omega_j\right|_{\lambda=const} =
- {\gamma_{j\alpha}\over
2\sqrt{\lambda-\lambda_\alpha}}{\d\lambda_\alpha\over\d t_k} + {\rm regular}
\ee
Then, using (\ref{yl}), (\ref{deryl}) and (\ref{difyv}) together with
\be
d\Omega_j = {\gamma_{j\alpha}\over
2\sqrt{\lambda-\lambda_\alpha}}\ d\lambda + \dots
\ee
and the relation, following from (\ref{omes}), (\ref{deryl})
\be
d\Omega_k = {\d\over\d t_k}dS  =
- {\Gamma_\alpha d\lambda\over
2\sqrt{\lambda-\lambda_\alpha}}\ {\d\lambda_\alpha\over\d t_k}
+ {\rm regular}
\ee
one gets for (\ref{3mero})
\be
{\d^3\F\over\d t_k\d t_n\d t_m} =
\sum \res_{\lambda_\alpha} \left(d\Omega_k {\d \Omega_n\over \d t_m}\right)
= - \sum \res_{\lambda_\alpha} \left(d\Omega_k{\gamma_{n\alpha}\over
2\sqrt{\lambda-\lambda_\alpha}}{\d\lambda_\alpha\over\d t_m}\right) =
\\
= - \sum \res_{\lambda_\alpha} \left(d\Omega_k{d\Omega_n\over d\lambda}
{\d\lambda_\alpha\over\d t_m}\right) =
\sum \res_{\lambda_\alpha} \left({d\Omega_kd\Omega_nd\Omega_m
\over d\lambda dy}\right)
\ee
In a similar way, one proves the residue formula for the mixed derivatives, so
that we finally conclude
\be
\label{resgen}
{\d^3 \F\over \d T_I\d T_J\d T_K} = \sum_{\lambda_\alpha}
\res_{\lambda_\alpha}\left(dH_IdH_JdH_K\over d\lambda dy\right) =
\sum_{\lambda_\alpha}\res_{\lambda_\alpha}
\left({\phi_I\phi_J\phi_K\over {d\lambda /dy}}dy\right) =\\=
\sum_{\lambda_\alpha}\Gamma_\alpha^2
\phi_I(\lambda_\alpha)\phi_J(\lambda_\alpha)
\phi_K(\lambda_\alpha)=\sum_{\lambda_{\alpha}}{\hat H_I(\lambda_{\alpha})
\hat H_J(\lambda_{\alpha})\hat H_K(\lambda_{\alpha})\over
\prod_{\beta\ne\alpha}(\lambda_{\alpha}-\lambda_{\beta})^2}
\ee
for the whole set $\{ T_I \} = \{ t_k, t_0, S_i \}$ and
$\{ dH_I \} = \{ d\Omega_k, d\Omega_0, d\omega_i \}$. In formula
(\ref{resgen}) we have introduced the meromorphic functions
\be
\label{phi}
\phi_I(\lambda) = {dH_I\over dy} = {\hat H_I(\lambda)\over R'(\lambda)}
\ee
for any (meromorphic or holomorphic) differential $dH_I = \hat
H_I(\lambda){d\lambda\over y}$ and
$R(\lambda) = W'(\lambda)^2 + f(\lambda)$. The derivation of the residue
formula for the set of parameters including $t_0$ corresponding to the
third-kind Abelian differential (\ref{bipole}) can be performed in a similar
way.

\section{Proof of WDVV
\label{ss:proof}}

Having the residue formula (\ref{resgen}), the proof of the WDVV equations
(\ref{WDVV}) is reduced to solving the system of linear equations
\cite{BMRWZ,Msusy,Mtmf}, which requires only fulfilling the two conditions:
\begin{itemize}

\item The ``matching" condition
\be
\label{matching}
\#(I)=\#(\alpha)
\ee
and

\item nondegeneracy of the matrix built from (\ref{phi}):
\be
\label{det}
\det_{I\alpha}\| \phi_{I}(\lambda_\alpha)\| \neq 0
\ee
\end{itemize}

Under these conditions, the structure constants
$C_{IJ}^K$ of the associative algebra
\be\label{CC}
\left(C_I\right)^K_L \left(C_J\right)^L_M = \left(C_J\right)^K_L
\left(C_I\right)^L_M
\\
(C_I)_J^K \equiv C_{IJ}^K
\ee
responsible for the
WDVV equations can be found
from the system of {\em linear equations}
\be
\label{eqc}
\phi_I(\lambda_\alpha)\phi_J(\lambda_\alpha) =\sum_K
C^K_{IJ}\phi_K(\lambda_\alpha), \ \ \ \ \ \ \ \forall\ \lambda_\alpha
\ee
with the solution
\be
\label{litc}
C^K_{IJ} = \sum_\alpha
\phi_I(\lambda_\alpha)\phi_J(\lambda_\alpha)
\left(\phi_K(\lambda_\alpha)\right)^{-1}
\ee
To make it as general, as in \cite{MMM,MMMlong,forms}, one may consider an
associative isomorphic algebra (again $\forall\ \lambda_\alpha$)
\be
\label{eqcxi}
\phi_I(\lambda_\alpha)\phi_J(\lambda_\alpha) =\sum_K
C^K_{IJ}(\xi)\phi_K(\lambda_\alpha)\cdot\xi(\lambda_\alpha)
\ee
which instead of (\ref{litc}) leads to
\be
\label{litcxi}
C^K_{IJ}(\xi) = \sum_\alpha
{\phi_I(\lambda_\alpha)\phi_J(\lambda_\alpha)\over\xi(\lambda_\alpha)}
\left(\phi_K(\lambda_\alpha)\right)^{-1}
\ee
The rest of the proof is consistency of this algebra with relation
\be
\label{feta}
{\cal F}_{IJK} = \sum_L C_{IJ}^L(\xi)\eta_{KL}(\xi)
\ee
with
\be
\label{metric}
\eta_{KL}(\xi) = \sum_A \xi_A {\cal F}_{KLA}
\ee
expressing structure constants in terms of the third derivatives and, thus,
leading to (\ref{WDVV}). It is easy to see that (\ref{feta}) is satisfied
if ${\cal F}_{KLA}$ are given by residue formula (\ref{resgen}).

Indeed, requiring {\em only} matching $\#(\alpha)=\#(I)$, one gets
\be
\sum_K C_{IJ}^K(\xi)\eta_{KL}(\xi) =
\sum_{K,\alpha,\beta}
{\phi_I(\lambda_\alpha)\phi_J(\lambda_\alpha)\over\xi(\lambda_\alpha)}\cdot
\left(\phi_K(\lambda_\alpha)\right)^{-1}\cdot\phi_K(\lambda_\beta)
\phi_L(\lambda_\beta)\xi(\lambda_\beta)\Gamma_\beta
\ee
and finally
\be
\sum_K C_{IJ}^K(\xi)\eta_{KL}(\xi) =
\sum_\alpha
{\phi_I(\lambda_\alpha)\phi_J(\lambda_\alpha)\over\xi(\lambda_\alpha)}
\phi_L(\lambda_\alpha)\xi(\lambda_\alpha)\Gamma_\alpha =
\sum_\alpha \Gamma_\alpha
\phi_I(\lambda_\alpha)\phi_J(\lambda_\alpha)\phi_L(\lambda_\alpha)
 = {\cal F}_{IJL}
\ee
Hence, for the proof of (\ref{WDVV}) one has to adjust the number of
parameters $\{ T_I\}$ according to (\ref{matching}). The number of critical
points $\#(\alpha) = 2n$ since $d\lambda=0$ for $y^2 = R(\lambda) = 0$.
Thus, one have to take a codimension one subspace in the space of all
parameters $\{ T_I\}$, a natural choice will be to fix the eldest
coefficient of (\ref{mmpot'}). Then the total number of parameters
$\#(I)$, including the periods ${\bf S}$, residue $t_0$ and the rest of
the coefficients of the potential will be $g+1+n = (n-1)+1+n = 2n$, i.e.
exactly equal to $\#(\alpha)$.
In sect.~\ref{ss:expl} we present an explicit
check of the WDVV equations for this choice, using the expansion of
free energy computed in \cite{CIV,IM3}.

Note that equations (\ref{eqcxi}) are basically equivalent to the
algebra of forms (or differentials) considered in \cite{forms,MMMlong}.
In this particular case one may take the basis of
1-differentials $d\omega_i$, $d\Omega_k$ with multiplication given
by usual (not wedge!) multiplication modulo
$dS=yd\lambda$. Then, one can either directly check that the algebra with
this multiplication is associative (similar to how it was done
in \cite{MMMlong,Luuk}), or, using hyperelliptic parameterization,
remove the factor ${d\lambda\over y}$ in order to reduce the
algebra to the ring of polynomials with multiplication modulo the
polynomial ideal $y^2=W'^2(\lambda)+f(\lambda)$, which is obviously
associative.

In the proof of the WDVV equations we used in this section, the
associativity (\ref{CC}) is absolutely evident, being associativity of
the usual multiplication, and the main point to
check was to derive (\ref{feta})-(\ref{metric}). When using instead
the algebra of differentials, the main
non-trivial point is to check its associativity, while the relations
(\ref{feta})-(\ref{metric}) appear as even more transparent than above
corollary of the residue formula.

\section{Explicit check of the WDVV equations
\label{ss:expl}}

To convince ourselves that the general proof
indeed works and to get some further insights, in this section
we consider the explicit check of the WDVV equations
(\ref{WDVV}) perturbatively. To do this, let us take the perturbative
expansion of the prepotential (\ref{FDV}) at small $S_i$'s (here we
take the symmetric choice of variables with $i=1,\dots,n=g+1$), keeping
the coefficients of the matrix model potential ${\bf t}$'s arbitrary,
and see if this perturbative expansion of $\F$ satisfies the WDVV
equations order by order. A general procedure of getting
such perturbative expansion was proposed in \cite{CIV}.

We are going to check here
only the simplest non-trivial case of cubic matrix model potential, which
is reformulated in terms of elliptic curve (\ref{dvc}) with four
branch points. According to (\ref{matching})
the corresponding solution to WDVV equations (\ref{WDVV}) should depend
exactly on four independent
variables, and we are choosing them to consist of two filling numbers ($S_1$
and $S_2$)
and two coefficients of the potential ($t_1$ and $t_2$). The perturbative
expansion for this case was constructed in \cite{CIV} up to the fifth order in
$S_i$'s and was later discussed in many other places
(see, e.g., \cite{DGKV,IM3}). It reads
\be
\label{pprep}
{\cal F}=-S_1W(\mu_1)-S_2W(\mu_2)+{1\over 2}S_1^2\log\left({S_1\over
\Delta}\right)+ {1\over 2}S_2^2\log\left({S_2\over \Delta}\right)
+2S_1S_2\log(\Delta)+\\+{1\over\Delta^3}\left({2\over 3}S_1^3-{2\over
3}S_2^3+5S_1S_2^2-5S_1^2S_2\right)+ {1\over\Delta^6}\left({8\over 3}S_1^4+{8\over 3}S_2^4
-{91\over 3}S_1S_2^3-{91\over 3}S_1^3S_2+59S_1^2S_2^2\right) +\\+
{1\over\Delta^9}\left({56\over 3}S_1^5-{56\over 3}S_2^5-{871\over 3}S_1^4S_2+{871\over
3}S_2^4S_1+{2636\over 3}S_1^3S_2^2-{2636\over 3}S_1^2S_2^3\right) +
O(S^6)
\ee
Here the matrix model potential (\ref{mmpot}) is fixed to be
$W(\lambda )=\lambda ^3/3 +t_2\lambda ^2/2+t_1\lambda $; and
$\mu_1$, $\mu_2$ are the roots of the equation $W'(\lambda )=0$, i.e.
\be
\mu_1=-{1\over 2}t_2+{1\over 2}\sqrt{t_2^2-4t_1}\\
\mu_2=-{1\over 2}t_2-
{1\over 2}\sqrt{t_2^2-4t_1}\\
\Delta\equiv\mu_1-\mu_2=\sqrt{t_2^2-4t_1}
\ee
The perturbative expansion of free energy (\ref{pprep})
is symmetric with respect to simultaneous transformation
$S_1\leftrightarrow -S_2$ and $t_2\leftrightarrow -t_2$. However, since the
whole prepotential but its linear part depends only on $t_2^2$, the
transformation of $t_2$ is only essential for the linear term.

Note that in the context of supersymmetric field theories such formulas
usually depend on an additional parameter $\Lambda_{QCD}$ associated
with the field theory scale \cite{CIV}. However, this parameter here
emerges only
as regularization (see footnote \ref{reg}) and generally can be omitted
within the matrix model approach. Moreover, since the terms depending on
$\Lambda_{QCD}$ are at most quadratic in $S_i$ and $t_i$ (they are
logarithmic or polynomial in $\Lambda_{QCD}$, with the latter proportional to
$W(\Lambda_{QCD})$), they do not contribute to the WDVV equations and just
renormalize the second derivatives of the prepotential (or period matrix of
the curve similar to (\ref{dvc}) playing the role of the set of effective
couplings in the field theory at low energies (cf. with \cite{SW})).

There is one subtlety with the perturbative prepotential (\ref{pprep})
-- when calculating it the authors of \cite{CIV} were interested only in
the terms depending on $S_i$. There could be, in principle, some terms
dependent only on $t_i$'s. This does not, however, happen in our case.
Indeed, these terms would survive in the limit of all $S_i=0$, i.e. when
$f(\lambda)=0$ in (\ref{dvc}) and $y=W'(\lambda)$ is just a polynomial.
Then from (\ref{v}) it follows that
$v_k={\partial {\cal F}\over\partial t_k}=0$ in this limit, hence
we miss at most linear in $t_i$ terms.

In contrast to the weak coupling expansion of the Seiberg-Witten theory
when logarithmic terms do satisfy the WDVV equations themselves {\em
exactly} (see \cite{MMM})\footnote{Note that expansion (\ref{pprep})
that goes over {\it positive} powers of
variables, $S_i$'s is rather similar not to the weak coupling expansion of
the Seiberg-Witten theory but to its
strong coupling expansion (see, for
example, \cite{DHoPho}). In the latter case, the WDVV equations
trivially hold due to the duality argument \cite{dWM}.
}, some technical complication with
perturbative check of whether the prepotential
(\ref{pprep}) satisfies the WDVV equations emerges due to different orders of
magnitude of different matrix elements of the matrix of third derivatives of
the prepotential (\ref{matrF}). Say, matrix $\left(\F_{t_2}\right)_{JK}$
has all
matrix elements except for $\left(\F_{t_2}\right)_{t_j,t_k}$
of the order $O(1)$, while
the latter ones of the order of $O(S_i)$. (We definitely put all $S_i$'s of
the same order of magnitude.)

Therefore, a careful calculation in every matrix element is required to determine
the order of the WDVV equations where the perturbative expansion contributes to.
In practice, one suffices to rescale all $S_i$'s with a scale
parameter $\Lambda$. Then, all non-trivial matrix elements of same matrix WDVV
equation (i.e. for concrete $IJK$) are of the same order in $\Lambda$ (choosing
other indices $IJK$ in (\ref{WDVV}) changes this order). Say, for the choice
$(I,J,K)=(t_2,S_1,t_1)$ the leading order of the equation(\ref{WDVV}) is $\Lambda^7$
and trivially
vanishes. It is contributed only by the linear and logarithmic (in $S_i$'s)
terms of (\ref{pprep}). The cubic terms of the prepotential contribute to
$\Lambda^8$-order, the quartic terms to $\Lambda^9$-order etc.
Using the program MAPLE, we have checked that the WDVV equations
(\ref{WDVV}) are satisfied with the prepotential
(\ref{pprep}) up to the fifth order in $S_i$.

Now, one can consider further non-trivial examples with
matrix model potentials being polynomials of higher degree. One technical
problem now is that $W(\mu_i)$ is getting more and more involved function of
$t_i$'s, since $\mu_i$'s become the roots of polynomial equations of increasing
order. Therefore, we lose just a principal possibility of analytic analyzing
$W(\mu_i)$ if $W(\lambda)$ is a polynomial of order 6 and higher.
Nevertheless, one
can still look at different limiting regimes with some of roots of
$W'(\lambda)$ approaching
close to each other.

\section{Conclusion}

In this paper we have proven that the (generalized) free energy of the
matrix model in planar limit satisfies the WDVV equations (\ref{WDVV}).
The proof is based on the residue formula for the quasiclassical
tau-functions introduced in \cite{KriW} which was derived above for
the whole basis of the first-kind or holomorphic (\ref{canoDV}),
second-kind (\ref{omes})
and third-kind (\ref{bipole}) meromorphic Abelian differentials.
The proof of the WDVV equations based on the residue formula requires
the matching condition (\ref{matching}) (or associativity of the
algebra of differentials), and the first nontrivial
example of our statement for the case of four variables we have checked
explicitly for its perturbative expansion (\ref{pprep}).

The matching condition (\ref{matching}) requires for
the quasiclassical tau-function defined by (\ref{dvc}) and (\ref{dvds})
to satisfy the WDVV equations (\ref{WDVV}) as a function of the
set of parameters involving necessarily the coefficients of the matrix model
potential and corresponding to the {\em meromorphic} differentials. In other
words, we cannot restrict ourselves for the set of only holomorphic
differentials, like it happens for the prepotentials of the Seiberg-Witten
models \cite{MMM}, associated as well in the sense of
sect.~\ref{ss:taumatr} to integrable systems \cite{GKMMM} (see
also \cite{Mbook} and references therein). This is similar to the case of
softly broken ${\cal N}=4$
Seiberg-Witten theory, where the WDVV equations are not satisfied by
the prepotential, being a function of the Seiberg-Witten periods corresponding
to the holomorphic differentials only \cite{MMMlong}. In that case one
should also necessarily add meromorphic differentials, but any physical
sense of the corresponding parameters, in contrast to the situation
considered in this paper, remains yet unclear.

Our statement can be generalized straightforwardly to the case of the
two-matrix model and the corresponding non-hyperelliptic curve \cite{KM}.
It was already suggested in \cite{KM} that the WDVV equations should hold
for the quasiclassical tau-function of the two-matrix model, whose
construction is based on the non-hyperelliptic curve, and similar to
(\ref{dvds}) meromorphic differential. One can use our line of reasoning
of sect.~\ref{ss:residue} and sect.~\ref{ss:proof} as a general proof
in the two-matrix case either, since our proof does not require any specific of
the curve (\ref{dvc}) like existence of the hyperelliptic
parameterization. The perturbative expansion of the multisupport
solution of the two-matrix model is not known yet (it can be computed, say,
using the diagrammatic expansion proposed in \cite{KM}) and it would be
interesting to calculate the free energy of the two matrix model explicitly
and repeat for this case the explicit calculation of sect.~\ref{ss:expl}.

There is another, very important direction where one can easily extend
the consideration of
the present paper. In \cite{CM} there was shown how one can introduce more
Whitham times, preserving the same number of moduli $S_i$. It is based on
using potential $W(\lambda)$ of higher degree and fixing some of the
coefficients of $f(\lambda)$ by the double point conditions. This procedure
allows one to make the number of Whitham times and moduli independent
in the WDVV equations. It would be
interesting to check if the WDVV equations are satisfied in this case, and
to construct the perturbative expansion for such a case.

Finally, let us point out that the fact that free energy of multi-support
solution of matrix model does satisfy the WDVV equations (\ref{WDVV}) can
be thought of as a first step to write down the generalization of the
(quasiclassical) Hirota equations which are known up to now only for the
one-support solutions. A link between the WDVV equations and (generalized)
Hirota relations \cite{BMRWZ,BraM} together with the statement of our paper
allows one to believe that the counterpart of the Hirota relations for nontrivial
quasiclassical tau-functions can be written in an explicit form.

\section*{Acknowledgements}

We are grateful to V.Kazakov, I.Krichever, A.Morozov and V.Rubtsov
for valuable discussions. The work
was partly supported by INTAS grant 00-561 (A.Mar, A.Mir. and D.V.), RFBR grants
01-01-00549 (L.Ch.), 02-02-16946 (A.Mar.), 01-01-00548 (A.Mir.) and
01-02-17488 (D.V.), by the Grants of Support of the Scientific
Schools 00-15-96046 (L.Ch.), 00-15-96566 (A.Mar) and 96-15-96798 (A.Mir.),
by Russian President's grant 00-15-99296 (D.V.),
by the Volkswagen-Stiftung (A.Mir. and D.V.)
and by the program ``Solitons" (L.Ch.).
A.Mar. would like to thank
the support of CNRS, Lab. de Physique Theorique of ENS, Paris, and IHES
where an essential part of this work has been done.


\begin{thebibliography}{12}

\bibitem{WDVV}
E.~Witten, Nucl. Phys. {\bf B340} (1990) 281;\\
R.~Dijkgraaf, H.~Verlinde and E.~Verlinde,
Nucl. Phys. {\bf B352} (1991) 59.
%
\bibitem{MMM}
A.~Marshakov, A.~Mironov and A.~Morozov,
Phys. Lett. {\bf B389} (1996) 43, hep-th/9607109.
%
\bibitem{forms}
A.~Marshakov, A.~Mironov and A.~Morozov,
Mod.Phys.Lett. {\bf A12} (1997) 773-787; hepth/9701014.
%
\bibitem{MMMlong}
A.~Marshakov, A.~Mironov and A.~Morozov,
Int.J.Mod.Phys. {\bf A15} (2000) 1157-1206;
hep-th/ 9701123.
%
\bibitem{Veselov}
A. Veselov, Phys.Lett. {\bf A261} (1999) 297-302, hep-th/9902142;
in the book ``Integrability: The Seiberg-Witten and Whitham Equations"
Ed. by H.W. Braden and I.M. Krichever. Gordon and Breach, 2000. Pp. 125-135;
hep-th/0105020.
%
\bibitem{Luuk}
R. Martini and P.K.H. Gragert, J. Nonlinear Math. Phys. {\bf 6} (1999), no. 1,
1-4, hep-th/9901166;\\
L.K. Hoevenaars, P.H.M. Kersten and R. Martini, Phys.Lett. {\bf B503} (2001)
189-196, hep-th/0012133;\\
L.K. Hoevenaars and R. Martini, Lett.Math.Phys. {\bf 57} (2001) 175-183,
hep-th/0102190.
%
\bibitem{BMRWZ}
A. Boyarsky, A. Marshakov, O. Ruchayskiy, P. Wiegmann and A. Zabrodin,
Phys. Lett. {\bf B515} (2001) 483-492; hep-th/0105260.
%
\bibitem{WDVVmore}
I.M. Krichever, Funct. Anal. Appl. {\bf 31} (1997) 25-39, hep-th/9611158;\\
K. Ito and S.-K. Yang, Phys.Lett. {\bf B415} (1997) 45-53, hep-th/9708017;
Phys.Lett. {\bf B433} (1998) 56-62, hep-th/9803126;\\
G. Bertoldi, M. Matone, Phys.Rev. {\bf D57} (1998) 6483-6485,
hep-th/9712109;\\
J.M. Isidro, Nucl.Phys. {\bf B539} (1999) 379-402, hep-th/9805051;\\
Y. Ohta, J.Math.Phys. {\bf 40} (1999) 4089-4098 (hep-th/9904121),
{\bf 41} (2000) 6042-6047 (hep-th/9905126);\\
J. van de Leur, nlin.SI/0004021;\\
H. Aratyn and J. van de Leur, hep-th/0104092.
%
\bibitem{CIV}
F. Cachazo, K. Intriligator and C. Vafa, Nucl.Phys. {\bf B603} (2001) 3-41;
hep-th/0103067;\\
F. Cachazo and C. Vafa, hep-th/0206017.
%
\bibitem{DV}
R.~ Dijkgraaf and C.~ Vafa,
hep-th/0206255;
hep-th/0207106;
hep-th/0208048.
%
\bibitem{SW}
N. Seiberg and E. Witten,
Nucl.Phys., {\bf B426} (1994) 19-52;
{\bf B431} (1994) 484-550.
%
\bibitem{Dub}
B. Dubrovin, {\it Geometry of 2-D topological field
     theories}, in: Integrable Systems and Quantum Groups
(Montecatini Terme, 1993), Lecture Notes in Math. {\bf 1620},
Springer, Berlin, 1996, 120-348, hep-th/9407018.
%
\bibitem{MirWDVV}
A. Mironov, In the book ``Integrability: The Seiberg-Witten and Whitham Equations"
Ed. by H.W. Braden and I.M. Krichever. Gordon and Breach, 2000;
hep-th/9903088.
%
\bibitem{Mtmf}
A.~Marshakov, ``On associativity equations,'' Theor.\ Math.\ Phys.\
{\bf 132}, 895 (2002) [Teor.\ Mat.\ Fiz.\ {\bf 132}, 3 (2002)]
arXiv:hep-th/0201267.
%
\bibitem{imo}
H. Itoyama and A. Morozov, "Experiments with the WDVV equations",
hep-th/0211259.
%
\bibitem{KriW}I.~Krichever,
Commun. Pure. Appl. Math. {\bf 47} (1992) 437, hep-th/9205110.
%
\bibitem{David}
F.~David,
Phys.Lett. B302 (1993) 403-410, hep-th/9212106;\\
G.~Bonnet, F.~David, B.~Eynard,
J.Phys. {\bf A33} (2000) 6739-6768, cond-mat/0003324.
%
\bibitem{CM}
 L.~Chekhov and A.~Mironov,
hep-th/0209085.
%
\bibitem{KM}
V.Kazakov and A.Marshakov, hep-th/0211236.
%
\bibitem{DV1}
N.~Dorey, T.~J.~Hollowood, S.~Prem Kumar and A.~Sinkovics, hep-th/0209089,
hep-th/0209099;\\
D. Berenstein, hep-th/0210183;\\
A. Gorsky, hep-th/0210281;\\
N.~Dorey, T.~J.~Hollowood and S.~P.~Kumar, hep-th/0210239;\\
R. Dijkgraaf, M.T. Grisaru, C.S. Lam, C. Vafa and D. Zanon,
hep-th/0211017;\\
S.G. Naculich, H.J. Schnitzer and N. Wyllard, hep-th/0211123,
hep-th/0211254;\\
Bo Feng, hep-th/0211202, hep-th/0212010;\\
Bo Feng and Y.-H. He, hep-th/0211234;\\
F.~Cachazo, M.~R.~Douglas, N.~Seiberg and E.~Witten,
hep-th/0211170;\\
F.~Cachazo, N.~Seiberg and E.~Witten, hep-th/0301006.
%
\bibitem{DV2}
G. Bonelli, hep-th/0209225;\\
H. Fuji and Y. Ookouchi, hep-th/0210148;\\
R. Argurio, V.L. Campos, G. Ferretti and R. Heise, hep-th/0210291,
hep-th/0211249;\\
J. McGreevy, hep-th/0211009;\\
H. Suzuki, hep-th/0211052, hep-th/0212121;\\
I. Bena and R. Roiban, hep-th/0211075;\\
Y.~Demasure, R.~A.~Janik, hep-th/0211082;\\
R. Gopakumar, hep-th/0211100;\\
I.~Bena, R.~Roiban and R.~Tatar, hep-th/0211271;\\
Y. Tachikawa, hep-th/0211274, hep-th/0211189;\\
Y. Ookouchi, hep-th/0211287;\\
S. Ashok, R. Corrado, N. Halmagyi, K. Kennaway and C. Romelsberger,
hep-th/0211291;\\
K.~Ohta, hep-th/0212025;\\
R.A. Janik and N.A. Obers, hep-th/0212069;\\
S. Seki, hep-th/0212079;\\
C. Hofman, hep-th/0212095.
%
\bibitem{Migdal}
A.~A.~Migdal,
Phys.\ Rept.\  {\bf 102}, 199 (1983).
%
\bibitem{RG}
A. Gorsky, A. Marshakov, A. Mironov and A. Morozov,
Nucl.Phys., {\bf B527} (1998) 690-716, hep-th/9802004.
%
\bibitem{Fay}
J. Fay, {\sl Theta-functions on Riemann surfaces,}
Lect. Notes Math. {\bf 352}, Springer, N.Y. 1973.
%
\bibitem{Kripri}
I. Krichever, private communication.
%
\bibitem{Msusy}
A. Marshakov, hep-th/0108023.
%
\bibitem{GKMMM}
A. Gorsky, I. Krichever, A. Marshakov, A. Mironov and A. Morozov,
Phys.Lett., {\bf B355} (1995) 466-477, hep-th/9505035.
%
\bibitem{Mbook}
A. Marshakov, {\sl Seiberg-Witten Theory and Integrable Systems},
World Scientific, 1999.
%
\bibitem{DHoPho}
E. D'Hoker and D.H. Phong, Phys.Lett. {\bf B397} (1997) 94-103,
hep-th/9701055.
%
\bibitem{dWM}
B. de Wit and A. Marshakov, hep-th/0105289.
%
\bibitem{IM3}
H. Itoyama and A. Morozov, hep-th/0212032.
%
\bibitem{DGKV}
R. Dijkgraaf, S. Gukov, V. Kazakov and C. Vafa, hep-th/0210238;\\
A. Klemm, M. Marino and S. Theisen, hep-th/0211216;\\
H. Itoyama and A. Morozov, hep-th/0211245.
%
\bibitem{BraM}
H.W. Braden and A. Marshakov, Phys. Lett.
{\bf B541} (2002) 376-383, hep-th/0205308.
%
\end{thebibliography}
\end{document}